\documentclass[prb,twocolumn,floatfix,longbibliography,superscriptaddress]{revtex4-2}
\usepackage[english]{babel}
\usepackage[utf8]{inputenc}
\usepackage[T1]{fontenc}
\usepackage{amsmath}
\usepackage{mathtools,amssymb}
\usepackage{graphicx}
\usepackage{bm}
\usepackage{gensymb}
\usepackage{color}
\usepackage{hyperref}
\usepackage{upgreek}
\usepackage{scalerel}
\usepackage{pgffor}
\usepackage{pdfpages}
\usepackage{float}
\usepackage{appendix}
\usepackage{physics} 
\usepackage{lscape}   
\usepackage{silence}
\usepackage{soul,xcolor} 
\setstcolor{red} 
\usepackage{comment}  
\makeatletter
\AtBeginDocument{\let\LS@rot\@undefined}
\makeatother

\begin{document}

\title{Tunneling conductance in superconducting junctions with $p$-wave unconventional magnets breaking time-reversal symmetry}

\author{Yuri Fukaya}
 \email[ ]{ yuri.fukaya@ec.okayama-u.ac.jp}
\affiliation{Faculty of Environmental Life, Natural Science and Technology, Okayama University, 700-8530 Okayama, Japan}

\author{Keiji Yada}
\email[]{ yada.keiji.b8@f.mail.nagoya-u.ac.jp}
\affiliation{Department of Applied Physics, Nagoya University, 464-8603 Nagoya, Japan}

\author{Yukio Tanaka}
 \email[ ]{ ytanaka@nuap.nagoya-u.ac.jp}
\affiliation{Department of Applied Physics, Nagoya University, 464-8603 Nagoya, Japan}
\affiliation{Research Center for Crystalline Materials Engineering, Nagoya University, 464-8603 Nagoya, Japan}

\date{\today}
\begin{abstract}
A new type of magnet called $p$-wave unconventional magnet is proposed, stimulated by the discovery of altermagnet.
We study the tunneling conductance of $p$-wave unconventional magnet/superconductor junctions by adopting the effective Hamiltonian of $p$-wave unconventional magnets with time-reversal symmetry breaking, suggested in Ref [arXiv: 2309.01607 (2024)]. 
The tunneling conductance shows an asymmetric behavior with respect to bias voltage in the helical $p$-wave superconductor junctions.
It is caused by the missing of helical edge states contributing to the charge conductance owing to the momentum-dependent spin-split feature of the Fermi surface in $p$-wave unconventional magnets. 
In chiral $d$ and $p$-wave superconductor junctions, the resulting spin-resolved tunneling conductance takes a different value for spin sectors due to the time-reversal symmetry breaking in superconductors.
Our results qualitatively reproduce the results based on the simplified Hamiltonian in Ref [J.\ Phys.\ Soc.\ Jpn.\ \textbf{93}, 114703 (2024)], where only the odd function of the exchange coupling of $p$-wave unconventional magnets is taken into account, which gives the shift of the Fermi surface and preserves the time-reversal symmetry similar to the spin-orbit coupling. 
\end{abstract}
  
\maketitle

\section{Introduction}
\label{section0}

Charge transport of superconducting junctions is a fundamental problem of superconductivity and mesoscopic physics due to the presence of Andreev reflection~\cite{Andreev1964,Blonder_and_Tinkham}.
Andreev reflection is a unique process in inhomogeneous superconductors (SCs), like superconducting junctions, where an injected electron (hole) is reflected as a hole (electron) at the interface, and it can generate the inner gap states called Andreev bound states \cite{Mcmillan1968,Kashiwaya_2000, Sauls2018}.
In unconventional SCs, when a quasiparticle feels a different phase of the pair potential, we can expect the inner gap states localized near the surface of SCs called surface Andreev bound states (SABSs)\cite{Buchholtz1981,Hara1986,HuPRL1994,Kashiwaya_2000} which crucially influences the charge transport of unconventional SC junctions \cite{Kashiwaya_2000,Lofwander_2001}. 
It is known that spin-singlet $d$-wave SCs or spin-triplet $p$-wave SCs with line nodes lead to a zero-bias conductance peak (ZBCP) in normal metal/SC junctions~\cite{Tanaka95,Tanaka96,Kashiwaya_2000}, and it is caused by the existence of the flat bands of SABSs at zero-energy~\cite{SchnyderPRB2011,YadaPRB2011,BrydonPRB2011,SatoPRB2011}. 
The physical origin of ZBCP is the perfect Andreev reflection with probability unity \cite{Tanaka95,Kashiwaya_2000}, and this process is equivalent to the tunneling via Majorana zero mode in topological superconductors \cite{Bolech,tanaka12,tanaka2024theory}. 
ZBCP has been observed experimentally in cuprates~\cite{KashiwayaPRB1995,LesueurPhysC1992,AlffPRBR1997,IguchiPRBR2000,CovingtonPRL1997,WeiPRL1008,BiswasPRL2002,BouscherJPCM2020}, Sr$_2$RuO$_4$~\cite{KashiwayaPRL2011}, heavy fermion CeCoIn$_5$~\cite{RourkePRL2005}, PuCoGa$_5$~\cite{DagheroNatCommun2012}, UBe$_{13}$~\cite{WaltiPRL2000}, UTe$_2$~\cite{Yoonnpj2024}, PrOs$_4$Sb$_{12}$~\cite{TurelJPSJ2008}, surface superconductivity AuSn$_4$~\cite{ZhuNatCommun2023}, CaAg$_{1-x}$Pd$_x$P~\cite{YanoNatCommun2023},  Cu$_x$Bi$_2$Se$_3$~\cite{SasakiPRL2011}, La$_3$Ni$_2$O$_7$~\cite{LiuSCPMA2025}, and RbCr$_3$As$_3$~\cite{LiuPRB2019}. 
Also, there have been studies about the charge transport of normal metal/unconventional SC junctions where the SABSs have dispersions like chiral $p$-wave~\cite{Yamashiro97,Honerkamp1998}
chiral $d$-wave~\cite{Kashiwaya2014} and helical $p$-wave~\cite{Iniotakis2007,TanakaNagaosaBalatsky}. 
In the case of dispersive SABS, the Andreev reflection and resulting conductance are sensitive to the Fermi surface of both the normal metal and SCs \cite{Sengupta}.
\par
Besides normal metal/SC junctions, there are many works about ferromagnet/SC junctions up to now \cite{Sarma_RevModPhys_2004,Linder_2015,Eschrig2015}. 
Even if we confine the discussions in the ballistic junctions, there were several theoretical studies up to now. 
In ferromagnet (FM)/conventional $s$-wave SC junctions, the Andreev reflection is suppressed by the exchange field, which induces the spin splitting of the Fermi surface~\cite{JongPRL1995}.
Theoretical studies in FM/SC junctions for the spin-singlet $d$-wave and the spin-triplet $p$-wave states were demonstrated~\cite{KashiwayaPRB1999,ZuticPRB1999,ZhuPRB2000,HiraiPRB2003}. 
The Andreev reflection and the height of a ZBCP in FM/$d$-wave SC junctions are suppressed by the spin splitting in FM.
While in FM/$p$-wave junctions, the effect of FM is determined by the relative direction between the exchange field of FM and the \textbf{d}-vector of the spin-triplet pairing~\cite{HiraiJPSJ2001}.

Recently, new type of magnets so-called ``altermagnets'' (AMs) was discovered~\cite{NakaNatCommun2019,Hayami19,NakaPRB2020,Hayami20,LiborSAv,MazinPNAS,MaNatcommun2021,LiborPRX22,MazinPRX22,Libor011028,landscape22,Brekke23,han2024SciAdv,fzhang2024,jiang2024,Helena2021,jungwirth2024H,krempasky2024,Osumi2024,Bai_review24,FukayaCayaoReview2025}.
As an extension of AMs, $p$-wave unconventional magnets (PUMs) were also proposed~\cite{BrekkePRL2024,hellenes2024P,yu2025oddparity} and observed experimentally in NiI$_2$~\cite{song2025electrical}.
There are many studies of AMs and PUMs exploring new phenomena in superconducting junctions with AMs and PUMs and relevant topics ~\cite{FukayaCayaoReview2025} 
e.g.\ Andreev reflection~\cite{Sun23,Papaj23,NagaePRBL2025}, Josephson effect~\cite{Ouassou23,Beenakker23,Cheng24,fukaya2024,zhao2025Lu,sharma2025,alipourzadeh2025,boruah2025,pal2025}, topological superconductivity~\cite{ZhuPRB2023,CCLiu1,CCLiu2,NagaePRB2025TSC,chatterjee2025interplay}, superconducting diode effect~\cite{chakraborty2024perfe,PhysRevB.110.014518,Banerjee24,sim2024}, the finite-momentum Cooper pair~\cite{zhang2024,MukasaJPSJ2024,hong2024,chakraborty2024}, and so on~\cite{chakraborty2024constraints,MaedaPRB2025,sukhachov2025coexistence,autieri2025,ramires2025,sun2025pseudoising,fu2025floquet}.
In Ref.\ \cite{maeda2024}, the tunneling conductance in PUM/SC junctions was calculated by using the simplified continuum model.
They found the suppression of Andreev bound states by the $p$-wave-like Fermi surface splitting in PUMs.
As a result, the zero-bias conductance shows a non-monotonic change as a function of the exchange energy of the PUM order.
However, in Ref.\ \cite{maeda2024}, they only considered the shift of the Fermi surfaces in the continuum model, and its Hamiltonian does not break the time-reversal symmetry, which is essentially equivalent to that of the persistent helix model ~\cite{BernevigPRL2006,KohdaPRB2012,LiuPRL2014,IkegayaPRB2015,PhysRevB.92.024510,YangPRB2017,AlidoustPRB2020,LeePRB2021,IkegayaPRB2021,AlidoustPRB2021,hellenes2024P}.
Because PUMs are caused by the noncollinear spin structure, the electronic structure of PUMs does not preserve the time-reversal symmetry~\cite{BrekkePRL2024,hellenes2024P}.
While, the energy spectrum in the momentum space satisfies the energy dispersion: $E(\bm{k},\sigma)=E(-\bm{k},-\sigma)$ with the momentum $\bm{k}$ and the spin $\sigma$, owing to $T\tau$-symmetry in Ref.~\cite{BrekkePRL2024} and $[C_2||t]$-symmetry in Ref.~\cite{hellenes2024P}.
In junctions with PUM, the spin structure near the surface/interface might affect transport properties.
By adopting the tight-binding model compared with the continuum model without the microscopic structure in the real space~\cite{maeda2024}, we can deal with not only the effective model of PUM~\cite{hellenes2024P} but also the spin structure of PUM at the interface. 
Thus, it is necessary to assess whether several theoretical predictions in an effective PUM model \cite{maeda2024} are justified or not by a model taking into account the time-reversal symmetry breaking.

In this paper, we investigate the tunneling conductance in superconducting junctions with PUMs. 
We adopt the model Hamiltonian of PUMs with time-reversal symmetry breaking suggested in Ref.\ \cite{hellenes2024P}.
For using this Hamiltonian, we adopt the tight-binding model to focus on the spin structure of PUMs at the interface, and we choose the junction along the $y$-direction in the low transparency limit.
For spin-singlet and spin-triplet SC junctions preserving time-reversal symmetry, although the range of the transverse momentum that contributes to tunneling conductance is restricted by the spin-split feature on the Fermi surface in PUMs, the line shape of the tunneling conductance is not qualitatively affected by the PUM order.
In the helical $p$-wave SC junction, the tunneling conductance is asymmetric for $eV$ with the bias voltage $V$ owing to the missing of helical edge states contributing to the conductance caused by the spin-split feature of the Fermi surface in PUM.
In chiral $d$ and chiral $p$-wave SC junctions, the spin-resolved tunneling conductance has a different value for spin sectors caused by the time-reversal symmetry breaking in superconductors.
Our calculations based on Ref.\ \cite{hellenes2024P}, taking account of the time-reversal symmetry breaking, qualitatively reproduce the behavior of the tunneling conductance obtained by 
the simplified model in Ref.\ \cite{maeda2024}. 

The construction of this paper is as follows.
In Sect.\ II, we show the model Hamiltonian and the procedure for the calculation of the tunneling conductance in PUM/SC junctions.
In Sect.\ III, we demonstrate the tunneling conductance and the momentum-resolved one for spin-singlet and spin-triplet pair potentials, including helical $p$-wave and chiral $d$ and $p$-wave superconductivity.
In Sect.\ IV, we conclude this work.

\begin{figure}[t!]
    \centering
    \includegraphics[width=8.5cm]{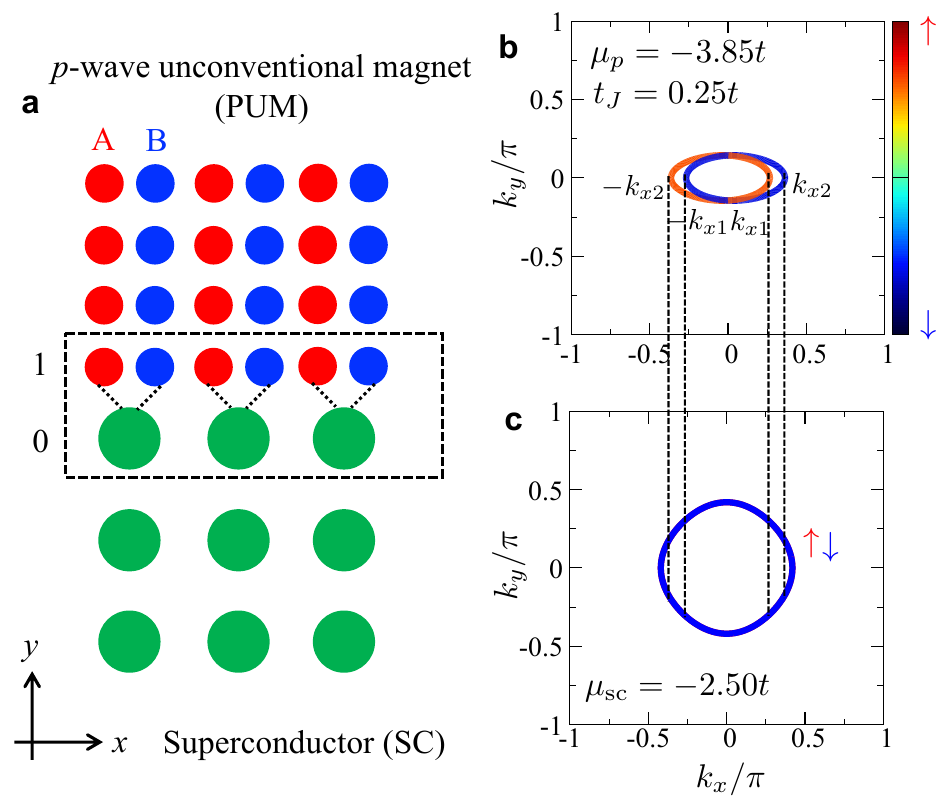}
    \caption{(a) Image of $p$-wave unconventional magnet (PUM)/superconductor (SC) junctions.
    We suppose the periodic boundary condition along the $x$-direction and semi-infinite systems along the $y$-direction.
    We consider two orbital (sublattice) degrees of freedom: orbital A (red) and B (blue) in PUMs and the single-orbital model in SCs (green). 
    To correspond to the $p_y$-wave magnet/SC junction with the interface along the $x$-direction, we consider a junction along the $y$-direction.
    Black dotted lines indicate the tunneling between A (red) and B (blue) of PUM and SC (green).
    0 and 1 indicate the indices in the Green's function of Sect.\ II.
    The Fermi surface of PUM and that in the normal state of SC are shown in (b) and (c). 
    (b) $\mu_{p}=-3.85t$ and $t_J=0.25t$, and (c) $\mu_\mathrm{sc}=-2.50t$.
    The black-dashed lines indicate the overlapping momenta $k_{x1}$ and $k_{x2}$ between PUM and SC.}
    \label{fig:image}
\end{figure}%

\section{Model and Methods}
\label{sec2}

In this section, we provide the model and the methods to calculate tunneling conductance. 
We adopt the tight-binding model, and focus on the single-orbital case in SCs and the effective two-sublattice model of PUM breaking time-reversal symmetry proposed in Ref.\ \cite{hellenes2024P}.
The Hamiltonian of SC in the normal state is described by the single-orbital model:
\begin{align}
    \hat{\mathcal{H}}_{0}&=\sum_{\bm{k}}\hat{C}^{\dagger}\hat{H}_{0}(\bm{k})\hat{C},
\end{align}%
with the creation and annihilation operators $\hat{C}^{\dagger}=[c^{\dagger}_{\uparrow},c^{\dagger}_{\downarrow}]$ and $\hat{C}$. 
$\hat{H}_{0}(\bm{k})$ is defined as
\begin{align}
    \hat{H}_{0}(\bm{k})=[-\mu_\mathrm{sc}-2t_1\cos{k_x}-2t_1\cos{k_y}]\hat{s}_{0},
\end{align}%
where $\mu_\mathrm{sc}=-2.5t$, $t_{1}=t$, and $\hat{s}_{0,x,y,z}$ are the chemical potential, hopping integral, and the Pauli matrices in the spin state, respectively.
In SCs, the Bogoliubov--de Gennes (BdG) Hamiltonian is given by
\begin{align}
    \hat{H}_\mathrm{BdG}=
    \begin{pmatrix}
        \hat{H}_{0}(\bm{k}) & \hat{\Delta}(\bm{k})\\
        \hat{\Delta}^{\dagger}(\bm{k}) & -\hat{H}^{*}_{0}(-\bm{k})
    \end{pmatrix},
\end{align}%
with the pair potential $\hat{\Delta}(\bm{k})$.
In the present study, we adopt both spin-singlet $\psi(\bm{k})$ and spin-triplet pair potential $\bm{d}(\bm{k})$~\cite{SigristUeda} with
\begin{align}
    \hat{\Delta}(\bm{k})=[\psi(\bm{k})+\bm{d}(\bm{k})\cdot\hat{\bm{s}}]i\hat{s}_{2}.
\end{align}%
Then, for the spin-singlet state, we focus on the $s$-wave:
\begin{align}
    \psi(\bm{k})=\Delta,
\end{align}%
$d_{x^2-y^2}$-wave:
\begin{align}
    \psi(\bm{k})=\Delta[\cos{k_x}-\cos{k_y}],
\end{align}%
and $d_{xy}$-wave superconductivity:
\begin{align}
    \psi(\bm{k})=2\Delta\sin{k_x}\sin{k_y}.
\end{align}%
For spin-triplet pairing, we choose the helical $p$-wave:
\begin{align}
    d_{x}(\bm{k})=\Delta\sin{k_y}, \hspace{2mm}d_{y}(\bm{k})=-\Delta\sin{k_x}, \hspace{2mm} d_{z}(\bm{k})=0,
\end{align}%
$p_x$-wave:
\begin{align}
    d_{x}(\bm{k})=0, \hspace{2mm}d_{y}(\bm{k})=0, \hspace{2mm} d_{z}(\bm{k})=\Delta\sin{k_x},
\end{align}%
and $p_y$-wave states:
\begin{align}
    d_{x}(\bm{k})=0, \hspace{2mm}d_{y}(\bm{k})=0, \hspace{2mm} d_{z}(\bm{k})=\Delta\sin{k_y}.
\end{align}%
We also consider the chiral $d$-wave
\begin{align}
    \psi(\bm{k})=\Delta[\alpha_{d1}(\cos{k_x}-\cos{k_y})+2i\alpha_{d2}\sin{k_x}\sin{k_y}],
\end{align}%
with $\alpha_{d1}=\alpha_{d2}=1$, and the chiral $p$-wave states
\begin{align}
    d_{x}(\bm{k})&=0, \hspace{2mm}d_{y}(\bm{k})=0,\notag \\
    d_{z}(\bm{k})&=\Delta[\alpha_{p1}\sin{k_x}+i\alpha_{p2}\sin{k_y}],
\end{align}%
with $\alpha_{p1}=\alpha_{p2}=1$.
We select the amplitude of the pair potential as $\Delta=0.001t$ in the following.
It is noted that we do not assume any magnetic order in SCs.
In PUM, we adopt the effective Hamiltonian proposed in Ref.\ \cite{hellenes2024P}:
\begin{align}
    \hat{\mathcal{H}}_\mathrm{PUM}&=\sum_{\bm{k}}\hat{B}^{\dagger}\hat{H}_\mathrm{PUM}(\bm{k})\hat{B},
\end{align}%
with the creation and annihilation operators $\hat{B}^{\dagger}=[b^{\dagger}_{A\uparrow},b^{\dagger}_{B\uparrow},b^{\dagger}_{A\downarrow},b^{\dagger}_{B\downarrow}]$ and $\hat{B}$.  
$\hat{H}_\mathrm{PUM}(\bm{k})$ is given by
\begin{align}
    &\hat{H}_\mathrm{PUM}(\bm{k})\notag\\
    &=-\mu_{p}\hat{s}_{0}\otimes\hat{\sigma}_{0}-2t_{p}\left[\cos\frac{k_x}{2}\hat{s}_{0}\otimes\hat{\sigma}_{1}+\cos{k_y}\hat{s}_{0}\otimes\hat{\sigma}_{0}\right]\notag\\
    &+2t_{J}\left[\sin\frac{k_x}{2}\hat{s}_{1}\otimes\hat{\sigma}_{2}+\cos{k_y}\hat{s}_{2}\otimes\hat{\sigma}_{3}\right],
\end{align}%
where $t_{p}=t$ is the nearest neighbor hopping integral in PUM, $t_J$ is the strength of the PUM order, and $\hat{\sigma}_{0,1,2,3}$ are the Pauli matrices in orbital (sublattice) space, respectively.
This PUM model has two orbital (sublattice) degrees of freedom: orbital (sublattice) $A$ and $B$.
For $t_{J}\ne 0$, $[C_2||t]$-symmetry is preserved, even though time-reversal symmetry is broken.
If $t_{J}\ne 0$, the electric structure holds the spin splitting of the PUM order; otherwise, a spin-degenerate Fermi surface is obtained.

\begin{figure}[t!]
    \centering
    \includegraphics[width=6cm]{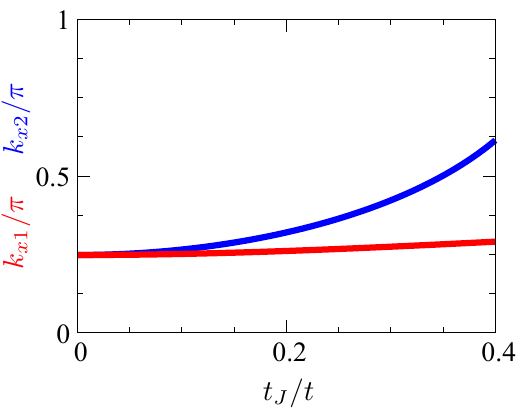}
    \caption{Values of $k_{x1}$ (red) and $k_{x2}$ (blue lines) as a function of the strength of the PUM order $t_J$. We choose the chemical potential as $\mu_{p}=-3.85t$.}
    \label{fig:kx1_kx2}
\end{figure}%

In the present study, we consider the junctions along the $y$-direction shown in Fig.~\ref{fig:image} (a) corresponding to $p_y$-wave UM/SC junctions along the $x$-direction.
Then, we suppose the periodic boundary condition along the $x$-direction and the semi-infinite systems along the $y$-direction, where the translational invariance is broken.
The Fermi surfaces on two sides of the junctions are chosen as $\mu_\mathrm{sc}=-2.5t$ and $\mu_{p}=-3.85t$ [Fig.~\ref{fig:image} (b) and Fig.~\ref{fig:image}(c)].
Since the tunneling conductance is determined by the overlapping area of Fermi surfaces between two sides of the junctions for for fixed $k_x$, we show how two Fermi surfaces of UPM and SC with $k_x$ in Fig.~\ref{fig:image} (b) and Fig.~\ref{fig:image} (c).
Only up-spin (down-spin) configuration on the Fermi surface in PUM appears for $-k_{x2} \le k_x\le -k_{x1}$ ($k_{x1} \le k_x\le k_{x2}$) shown in Fig.~\ref{fig:image} (b) and Fig.~\ref{fig:image} (c). 
These $k_{x1}$ and $k_{x2}$ changes by $t_J$ shown in Fig.~\ref{fig:kx1_kx2}.
Because $k_{x2}$ becomes larger for $t_J$, the Fermi surface of PUMs becomes larger with the increase of $t_J$.

\begin{figure*}[t!]
    \centering
    \includegraphics[width=14.5cm]{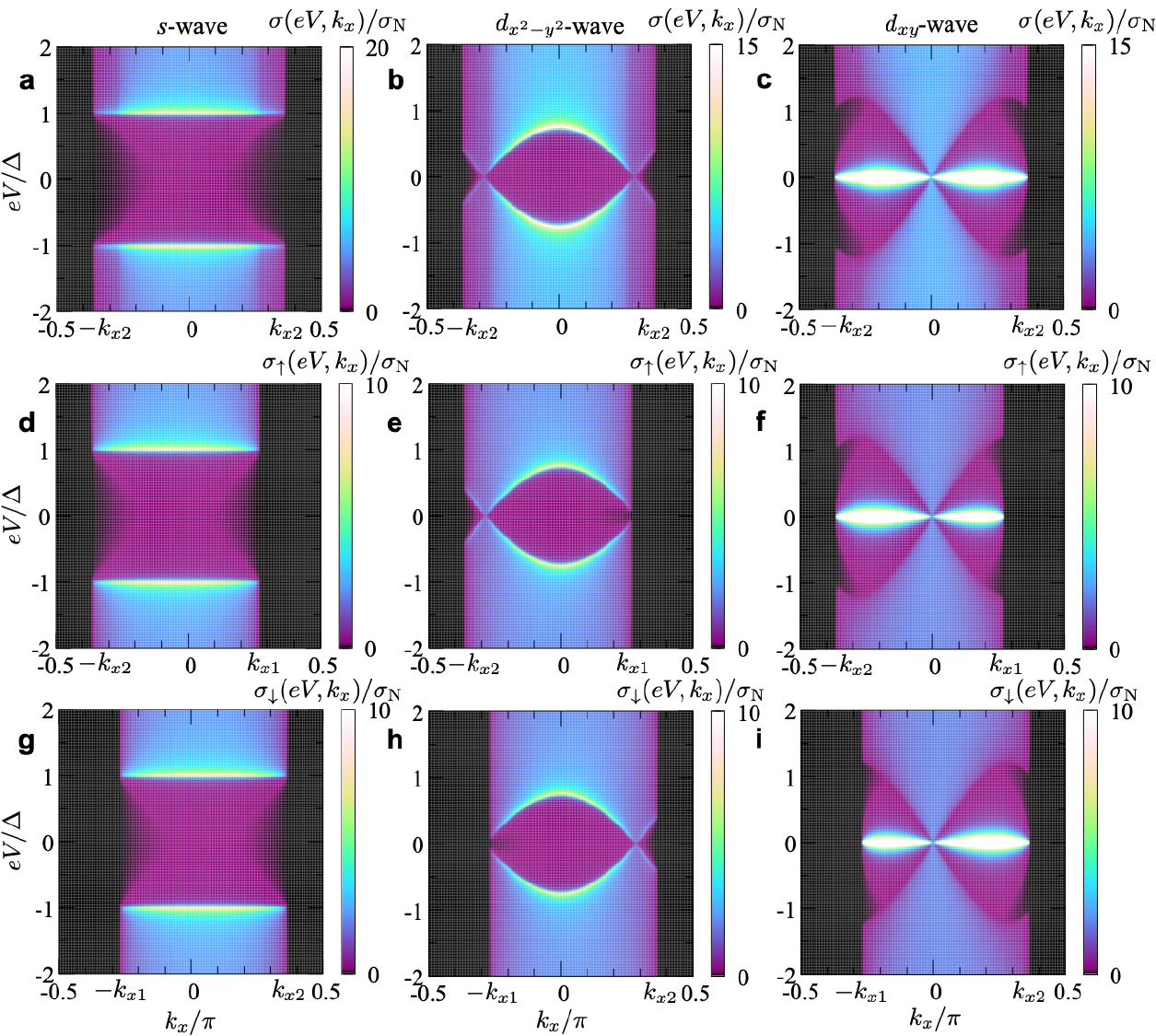}
    \caption{(a)(b)(c) Momentum-resolved tunneling conductance $\sigma(eV,k_x)$ for (a) spin-singlet $s$-wave, (b) $d_{x^2-y^2}$-wave, and (c) $d_{xy}$-wave SC junctions.
    Here, $\sigma_\mathrm{N}$ is the tunneling conductance in the normal state at $eV=0$.
    (d)(e)(f) Momentum-resolved tunneling conductance with up-spin $\sigma_{\uparrow}(eV,k_x)$ for (d) spin-singlet $s$-wave, (e) $d_{x^2-y^2}$-wave, and (f) $d_{xy}$-wave SC junctions.
    (g)(h)(i) Momentum-resolved tunneling conductance with down-spin $\sigma_{\downarrow}(eV,k_x)$ for (g) spin-singlet $s$-wave, (h) $d_{x^2-y^2}$-wave, and (i) $d_{xy}$-wave SC junctions.
    We choose the parameters as $t_J=0.25t$, $U_\mathrm{b}=5t$, $\Delta=0.001t$, and $\delta=0.01\Delta$.
    }
    \label{fig:k_resolved_singlet}
\end{figure*}%
\begin{figure*}[t!]
    \centering
    \includegraphics[width=12.5cm]{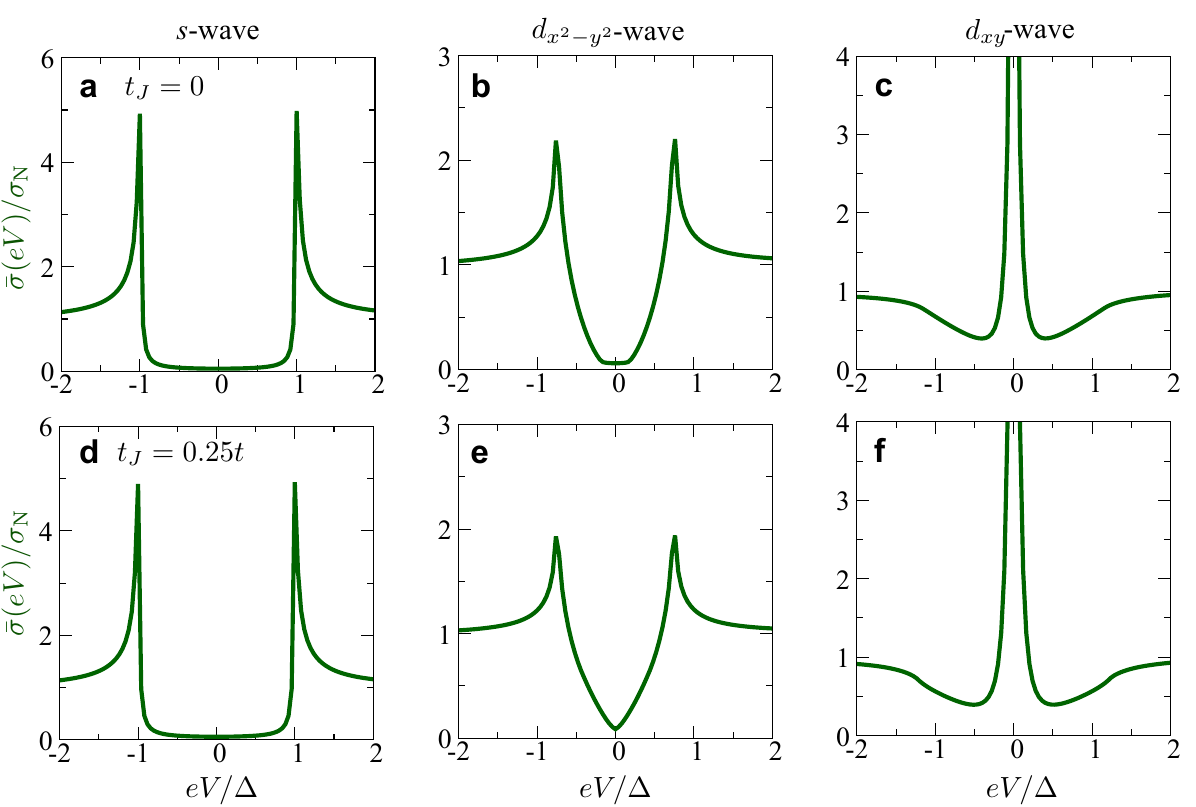}
    \caption{Tunneling conductance $\bar\sigma(eV)$ for (a)(d) spin-singlet $s$-wave, (b)(e) $d_{x^2-y^2}$-wave, and (c)(f) $d_{xy}$-wave SC junctions at (a)(b)(c) $t_J=0$ and (d)(e)(f) $t_J=0.25t$.
    It is normalized by its value in the normal state at $eV=0$: $\sigma_\mathrm{N}$.
    We choose the parameters as $U_\mathrm{b}=5t$, $\Delta=0.001t$, and $\delta=0.01\Delta$.}
    \label{fig:singlet}
\end{figure*}%
\begin{figure*}[t!]
    \centering
    \includegraphics[width=14.5cm]{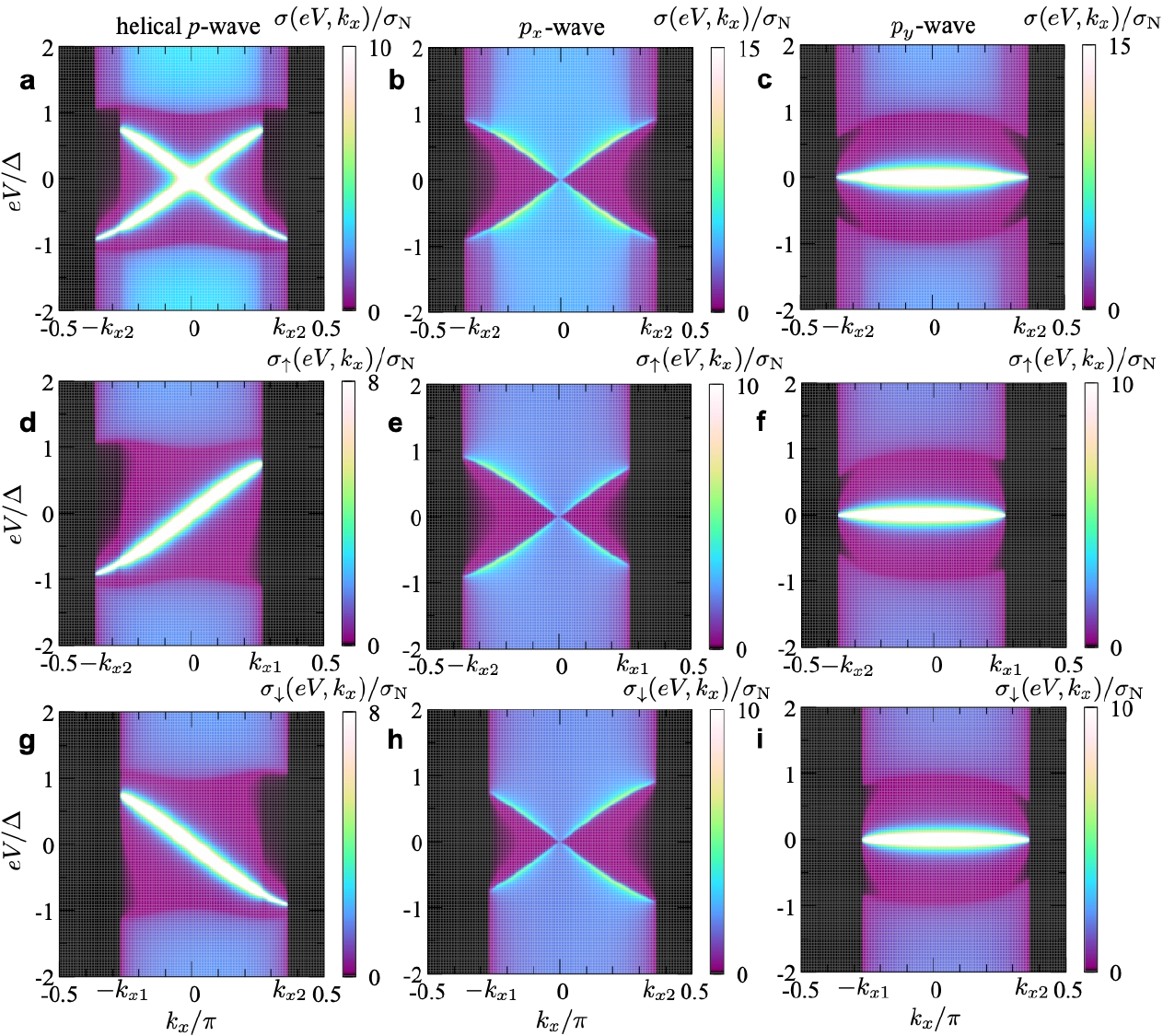}
    \caption{(a)(b)(c) Momentum-resolved tunneling conductance $\sigma(eV,k_x)$ for spin-triplet SC junctions. 
    (a)helical $p$-wave, (b) $p_{x}$-wave, and (c) $p_{y}$-wave SC junctions.
    Here, $\sigma_\mathrm{N}$ is the tunneling conductance in the normal state at $eV=0$.
    (d)(e)(f) Momentum-resolved tunneling conductance with up-spin $\sigma_{\uparrow}(eV,k_x)$ for (d)helical $p$-wave, (e) $p_{x}$-wave, and (f) $p_{y}$-wave SC junctions.
    (g)(h)(i) Momentum-resolved tunneling conductance with down-spin $\sigma_{\downarrow}(eV,k_x)$ for (d)helical $p$-wave, (e) $p_{x}$-wave, and (f) $p_{y}$-wave SC junctions.
    We choose the parameters as $t_J=0.25t$, $U_\mathrm{b}=5t$, $\Delta=0.001t$, and $\delta=0.01\Delta$.}
    \label{fig:k_resolved_triplet}
\end{figure*}%
\begin{figure*}[t!]
    \centering
    \includegraphics[width=12.5cm]{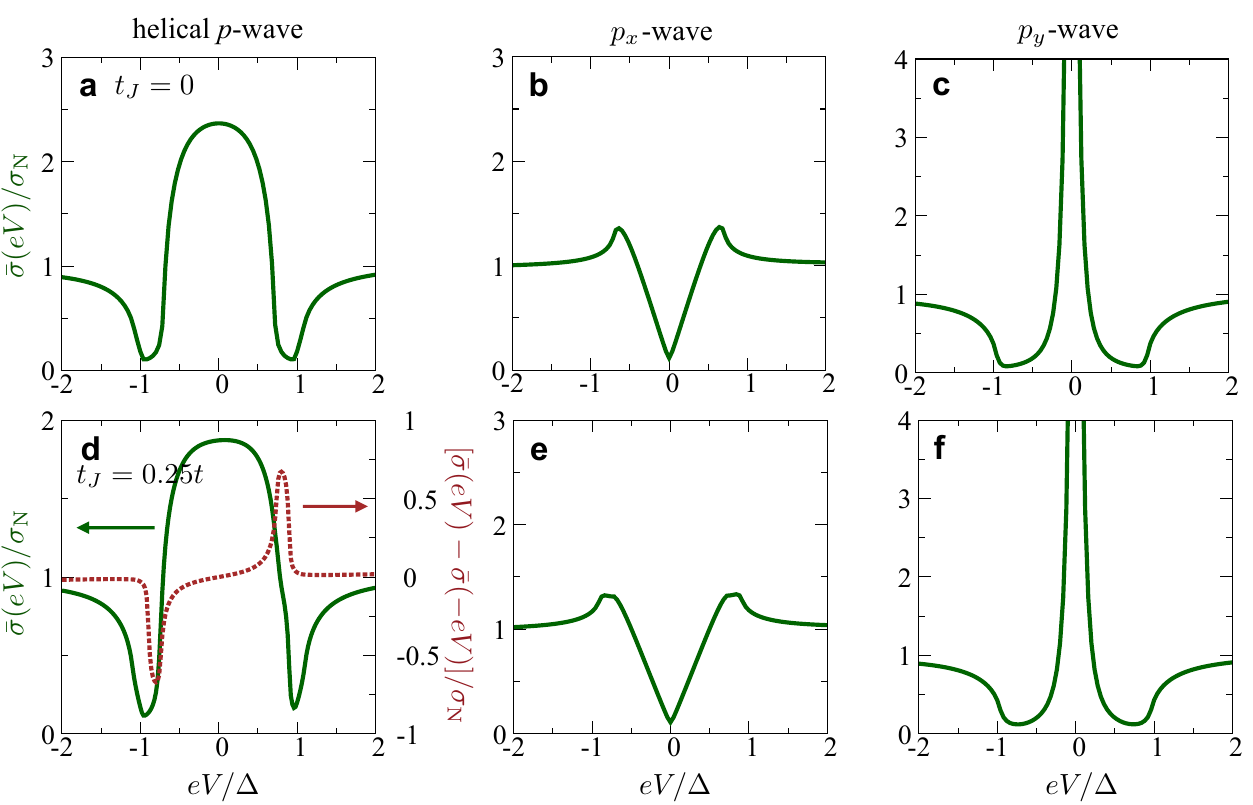}
    \caption{Tunneling conductance $\bar\sigma(eV)$ for spin-triplet SC junctions. 
    (a)(d) helical $p$-wave, (b)(e) $p_x$-wave, and (c)(f) $p_y$-wave SC junctions at (a)(b)(c) $t_J=0$ and (d)(e)(f) $t_J=0.25t$.
    In (d), we also plot $[\bar\sigma(eV)-\bar\sigma(eV)]/\sigma_\mathrm{N}$ to show the asymmetry of the conductance for $eV$.
    The tunneling conductance is normalized by its value in the normal state at $eV=0$: $\sigma_\mathrm{N}$.
    We choose the parameters as $U_\mathrm{b}=5t$, $\Delta=0.001t$, and $\delta=0.01\Delta$.}
    \label{fig:triplet}
\end{figure*}%
\begin{figure*}[t!]
    \centering
    \includegraphics[width=17.5cm]{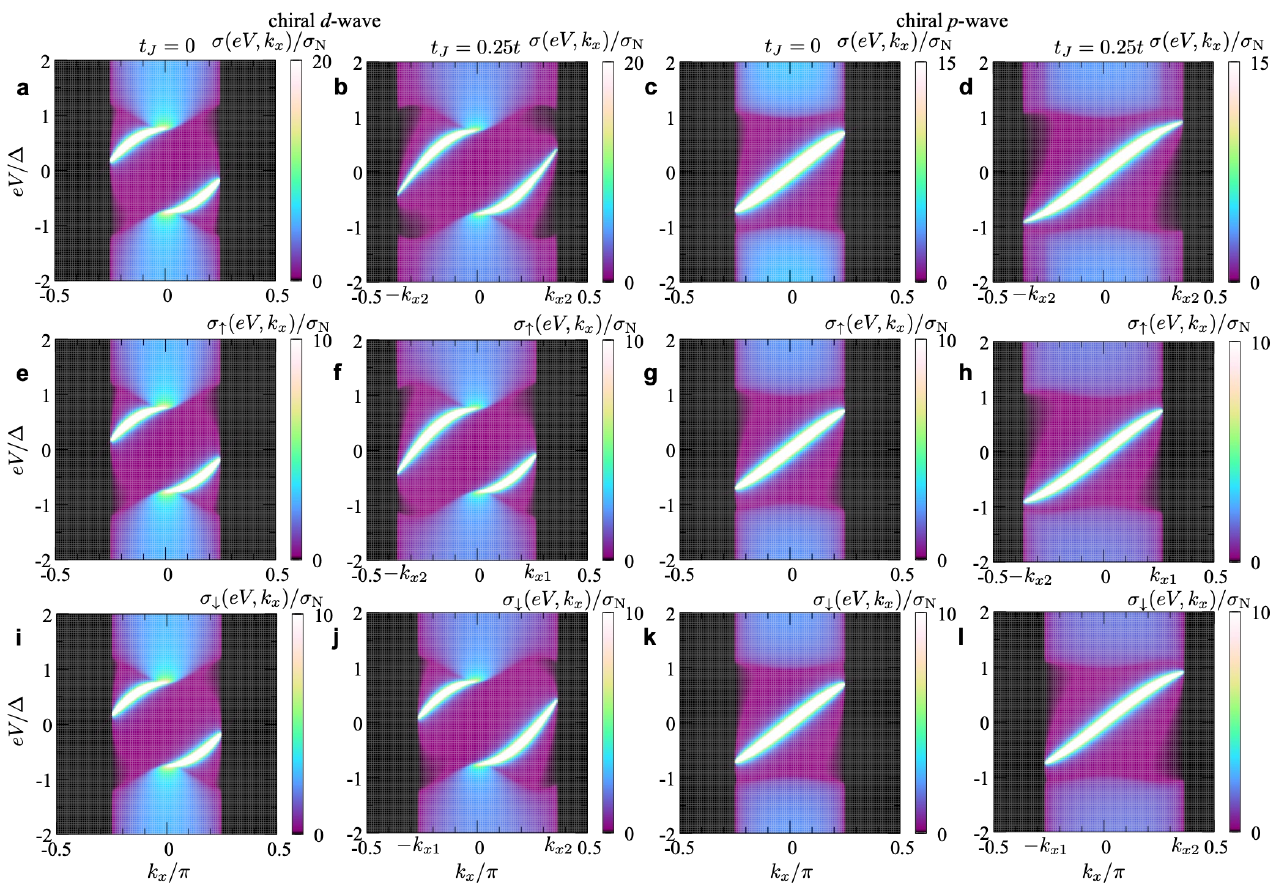}
    \caption{(a)(b)(c)(d) Momentum-resolved tunneling conductance $\sigma(eV,k_x)$ for (a)(b) chiral $d$-wave at (a) $t_J=0$ and (b) $t_J=0.25t$, and (c)(d) chiral $p$-wave SC junctions at (c) $t_J=0$ and (d) $t_J=0.25t$.
    Here, $\sigma_\mathrm{N}$ is the tunneling conductance in the normal state at $eV=0$.
    (e)(f)(g)(h) Momentum-resolved tunneling conductance with up-spin $\sigma_{\uparrow}(eV,k_x)$ for (e)(f) chiral $d$-wave at (e) $t_J=0$ and (f) $t_J=0.25t$, and (g)(h) chiral $p$-wave SC junctions at (g) $t_J=0$ and (h) $t_J=0.25t$.
    (i)(j)(k)(l) Momentum-resolved tunneling conductance with down-spin $\sigma_{\downarrow}(eV,k_x)$ for (i)(j) chiral $d$-wave at (i) $t_J=0$ and (j) $t_J=0.25t$, and (k)(l) chiral $p$-wave SC junctions at (k) $t_J=0$ and (l) $t_J=0.25t$. 
    We choose the parameters as $U_\mathrm{b}=5t$, $\Delta=0.001t$, and $\delta=0.01\Delta$.}
    \label{fig:k_resolved_chiral}
\end{figure*}%
\begin{figure}[t!]
    \centering
    \includegraphics[width=8.5cm]{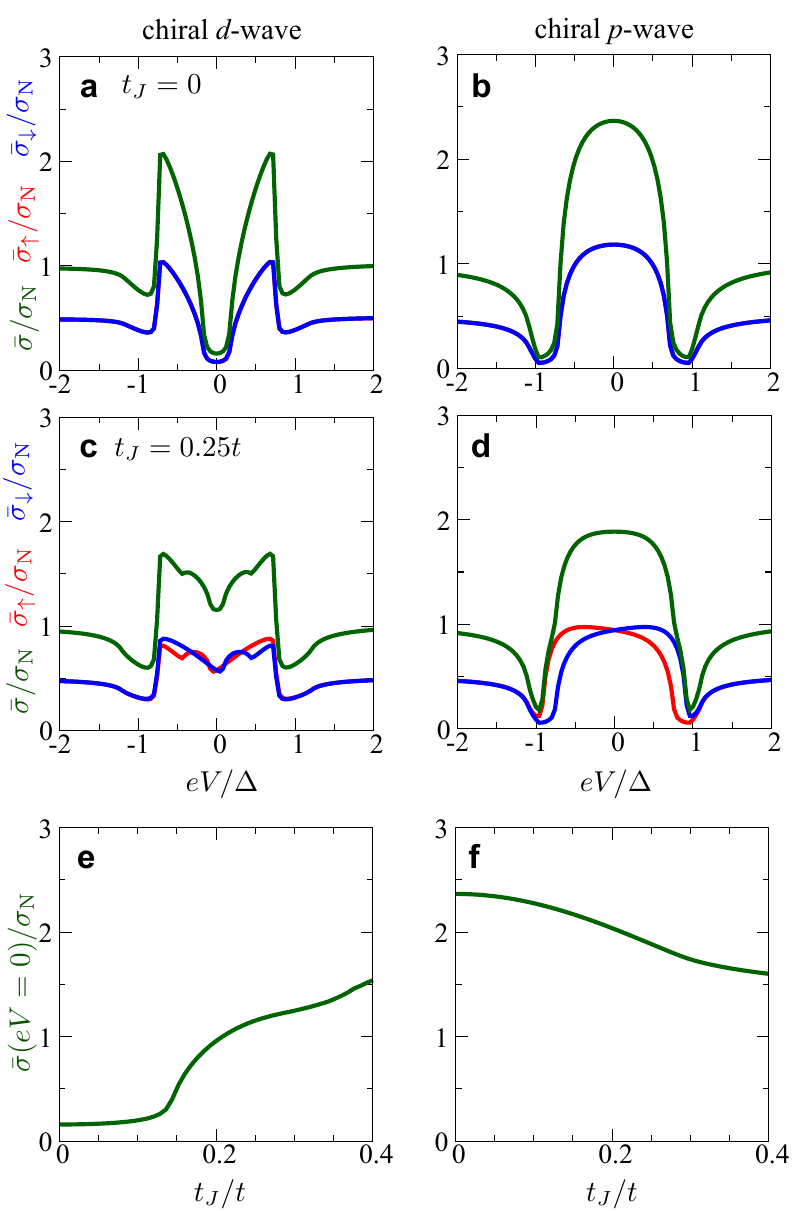}
    \caption{Tunneling conductance $\bar\sigma(eV)$ for (a)(c) chiral $d$-wave and (b)(d) chiral $p$-wave at (a)(b) $t_J=0$ and (c)(d) $t_J=0.25t$.
    This tunneling conductance is normalized by its value in the normal state at $eV=0$: $\sigma_\mathrm{N}$.
    Green, red and blue colors indicate $\bar\sigma(eV)/\sigma_\mathrm{N}$, $\bar\sigma_\uparrow(eV)/\sigma_\mathrm{N}$, and $\bar\sigma_\downarrow(eV)/\sigma_\mathrm{N}$, respectively.
    (e)(f) Tunneling conductance at $eV=0$ as a function of the strength of the PUM order $t_J$ for (e) chiral $d$-wave and (f) chiral $p$-wave SC junctions.
    We choose the parameters as $U_\mathrm{b}=5t$, $\Delta=0.001t$, and $\delta=0.01\Delta$.}
    \label{fig:chiral}
\end{figure}%

For the Fourier transformation along the $y$-direction, we obtain the local and the nearest neighbor (NN) hopping matrices. 
In SCs, the local term $\tilde{u}_\mathrm{SC}(k_x)$ is given by
\begin{align}
    \tilde{u}_\mathrm{SC}(k_x)&=
    \begin{pmatrix}
        \hat{u}_{0}(k_x) & \hat{u}_{\Delta}(k_x) \\
        \hat{u}^{\dagger}_{\Delta}(k_x) & -\hat{u}^{*}_{0}(-k_x)
    \end{pmatrix},
\end{align}%
\begin{align}
    \hat{u}_{0}(k_x)&=[-\mu_\mathrm{sc}-2t_{1}\cos{k_x}]\hat{s}_{0},
\end{align}%
and the NN hopping $\tilde{t}_\mathrm{SC}(k_x)$ is
\begin{align}
    \tilde{t}_\mathrm{SC}(k_x)=
    \begin{pmatrix}
        \hat{t}_{0}(k_x) & \hat{t}_{\Delta}(k_x) \\
        \bar{t}_{\Delta}(k_x) & -\hat{t}^{*}_{0}(-k_x)
    \end{pmatrix}
\end{align}%
\begin{align}
    \hat{t}_{0}&=-t\hat{s}_{0}.
\end{align}%
Here, $\hat{u}_{\Delta}(k_x)$ and $\hat{t}_{\Delta}(k_x)$ stand for the local and NN hopping terms for the pair potential. 
In PUMs, we obtain
\begin{align}
    \tilde{u}_\mathrm{PUM}(k_x)&=
    \begin{pmatrix}
        \hat{u}_\mathrm{PUM}(k_x) & 0 \\
        0 & -\hat{u}^{*}_\mathrm{PUM}(-k_x)
    \end{pmatrix},
\end{align}
\begin{align}
    \hat{u}_\mathrm{PUM}(k_x)&=-\mu_{p}\hat{s}_{0}\otimes\hat{\sigma}_{0}\\
    &-2t_{p}\cos\frac{k_x}{2}\hat{s}_{0}\otimes\hat{\sigma}_{1}+2t_{J}\sin\frac{k_x}{2}\hat{s}_{1}\otimes\hat{\sigma}_{2},\notag
\end{align}%
\begin{align}
    \tilde{t}_\mathrm{PUMs}(k_x)&=
    \begin{pmatrix}
        \hat{t}_\mathrm{PUM}(k_x) & 0 \\
        0 & -\hat{t}^{*}_\mathrm{PUM}(-k_x)
    \end{pmatrix},
\end{align}%
\begin{align}
    \hat{t}_\mathrm{PUM}(k_x)&=
    -t_{p}\hat{s}_{0}\otimes\hat{\sigma}_{0}+t_{J}\hat{s}_{2}\otimes\hat{\sigma}_{3}.
\end{align}%
Tunneling Hamiltonian $\hat{H}_{J}$ is described by using $\tilde{t}_\mathrm{J}$:
\begin{align}
    \tilde{t}_{\mathrm{J}}&=
    \hat{t}_{J}\otimes\hat{\tau}_{3},
\end{align}%
\begin{align}
    \hat{t}_{J}=
    \begin{pmatrix}
        -t_{1} & -t_{1} & 0 & 0 \\
        0 & 0 & -t_{1} & -t_{1} \\
    \end{pmatrix},
\end{align}
in the basis $[b^{\dagger}_{A\uparrow},b^{\dagger}_{B\uparrow},b^{\dagger}_{A\downarrow},b^{\dagger}_{B\downarrow}]$ and $[c_{\uparrow},c_{\downarrow}]$.
Here, $\hat{\tau}_{0,1,2,3}$ denote the Pauli matrices in Nambu space.

To calculate tunneling conductance, we adopt the recursive Green's function method~\cite{Umerski97}, which has been used in various problems of junctions 
\cite{Yada2014,Takagi2020}.
In general, the Green's function in the Nambu space is described by
\begin{align}
    \tilde{G}_{j_y,j'_y}(z,k_x)=
    \begin{pmatrix}
        \hat{G}_{j_y,j'_y}(z,k_x) & \hat{F}_{j_y,j'_y}(z,k_x)\\
        \bar{F}_{j_y,j'_y}(z,k_x) & \bar{G}_{j_y,j'_y}(z,k_x)
    \end{pmatrix},
\end{align}%
with $j_y$ and $j'_y$ the site indices along the $y$-direction. 
Here, $z=E\pm i\delta$ corresponds to the energy $E$
in the retarded and advanced parts of the Green's function 
with the infinitesimal value $\delta=0.01\Delta$.
Here, $\hat{F}(z,k_x)$ and $\bar{F}(z,k_x)$ denote the anomalous Green's function.
We see the tunneling conductance at the interface $j_x,j'_x=0,1$ shown in Fig.~\ref{fig:image} (a).
We obtain the local Green's function by the recursive Green's function method:
\begin{align}
    &\tilde{G}_{0,0}(z,k_x)\\
    &=[\tilde{G}^{-1}_{\mathrm{L},0}(z,k_x)-\tilde{t}_\mathrm{J}\tilde{G}_{\mathrm{R},1}(z,k_x)\tilde{t}^{\dagger}_\mathrm{J}]^{-1},\notag\\
    &\tilde{G}_{1,1}(z,k_x)\\
    &=[\tilde{G}^{-1}_{\mathrm{R},1}(z,k_x)-\tilde{t}^{\dagger}_\mathrm{J}\tilde{G}_{\mathrm{L},0}(z,k_x)\tilde{t}_\mathrm{J}]^{-1},\notag
\end{align}%
and the corresponding nonlocal part:
\begin{align}
    \tilde{G}_{0,1}(z,k_x)&=\tilde{G}_{\mathrm{L},0}(z,k_x)\tilde{t}_\mathrm{J}\tilde{G}_{1,1}(z,k_x),\\
    \tilde{G}_{1,0}(z,k_x)&=\tilde{G}_{\mathrm{R},1}(z,k_x)\tilde{t}^{\dagger}_\mathrm{J}\tilde{G}_{0,0}(z,k_x).
\end{align}%
Here, $\tilde{G}_{0,0}(z,k_x)$, $\tilde{G}_{1,1}(z,k_x)$, $\tilde{G}_{0,1}(z,k_x)$, and $\tilde{G}_{1,0}(z,k_x)$ are $4\times 4$, $8\times 8$, $4\times 8$, and $8\times 4$ matrices, respectively.
We first obtain the semi-infinite Green's function at the left and right-sides $\tilde{G}_\mathrm{L}(z,k_x)$ and $\tilde{G}_\mathrm{R}(z,k_x)$ in Ref.\ \cite{Umerski97}, and we add one layer at each side and we calculate the surface Green's function:
\begin{align}
    &\tilde{G}_{\mathrm{L},0}(z,k_x)\\
    &=[z-\tilde{u}_\mathrm{SC}(k_x)-\tilde{t}^{\dagger}_\mathrm{SC}(k_x)\tilde{G}_{\mathrm{L}}(z,k_x)\tilde{t}_\mathrm{SC}(k_x)]^{-1},\notag\\
    &\tilde{G}_{\mathrm{R},1}(z,k_x)\\
    &=[z-\tilde{u}_\mathrm{B}-\tilde{u}_\mathrm{PUM}(k_x)-\tilde{t}_\mathrm{PUM}(k_x)\tilde{G}_{\mathrm{R}}(z,k_x)\tilde{t}^{\dagger}_\mathrm{PUM}(k_x)]^{-1},\notag
\end{align}%
with the potential barrier at the interface $\tilde{u}_\mathrm{B}$ given by
\begin{align}
    \tilde{u}_\mathrm{B}=U_{b}\hat{s}_{0}\otimes\hat{\sigma}_{0}\otimes\hat{\tau}_{3},
\end{align}%
where $U_b=5t$ is the amplitude of the potential barrier at the interface.
Then, $\tilde{G}_{\mathrm{L},0}(z,k_x)$ and $\tilde{G}_{\mathrm{R},1}(z,k_x)$ are $4\times 4$ and $8\times 8$ matrices, respectively.
In the present study, we focus on the role of PUM order in superconducting junctions. It is noted that the transparency in the normal state does not become unity even without a potential barrier due to the mismatch of the two Fermi surfaces on the two sides of the junction. 
In the following, we consider the low transparent junctions by choosing the amplitude of the potential barrier as $U_{b}=5t$. 
We calculate the tunneling conductance based on the Lee-Fisher formula~\cite{Lee_Fisher,OhashiPRB2024}.
For each $E=eV$, the tunneling conductance is calculated by
\begin{align}
    \bar\sigma(eV)&=\frac{1}{2\pi}\int^{\pi}_{-\pi}dk_x\sigma(eV,k_x),
    \end{align}
\begin{align}
    &\sigma(eV,k_x)\\
    &=\mathrm{Tr'}[\tilde{g}_{1}(eV,k_x)+\tilde{g}_{2}(eV,k_x)-\tilde{g}_{3}(eV,k_x)-\tilde{g}_{4}(eV,k_x)],\notag
\end{align}%
with
\begin{align}
    \tilde{g}_{1}(E,k_x)&=\tilde{G}_{1,1}(E,k_x)\bar{t}^{\dagger}_\mathrm{J}\tilde{G}_{0,0}(E,k_x)\bar{t}_{J},\\
    \tilde{g}_{2}(E,k_x)&=\tilde{G}_{0,0}(E,k_x)\bar{t}_\mathrm{J}\tilde{G}_{1,1}(E,k_x)\bar{t}^{\dagger}_{J},\\
    \tilde{g}_{3}(E,k_x)&=\tilde{G}_{0,1}(E,k_x)\bar{t}^{\dagger}_\mathrm{J}\tilde{G}_{0,1}(E,k_x)\bar{t}^{\dagger}_{J},\\
    \tilde{g}_{4}(E,k_x)&=\tilde{G}_{1,0}(E,k_x)\bar{t}_\mathrm{J}\tilde{G}_{1,0}(E,k_x)\bar{t}_{J}.
\end{align}%
Then $\tilde{g}_{1}(E,k_x)$ and $\tilde{g}_{4}(E,k_x)$ are $8\times 8$, and $\tilde{g}_{2}(E,k_x)$ and $\tilde{g}_{3}(E,k_x)$ are $4\times 4$ matrices, respectively.
$\tilde{G}_{j_x,j'_x}(E,k_x)$ ($j_x,j'_x=0,1$) are defined as
\begin{align}
    \tilde{G}_{0,0}(E,k_x)&=-\frac{i}{2}[\tilde{G}^\mathrm{A}_{0,0}(E,k_x)-\tilde{G}^\mathrm{R}_{0,0}(E,k_x)],\\
    \tilde{G}_{1,1}(E,k_x)&=-\frac{i}{2}[\tilde{G}^\mathrm{A}_{1,1}(E,k_x)-\tilde{G}^\mathrm{R}_{1,1}(E,k_x)],\\
    \tilde{G}_{0,1}(E,k_x)&=-\frac{i}{2}[\tilde{G}^\mathrm{A}_{0,1}(E,k_x)-\tilde{G}^\mathrm{R}_{0,1}(E,k_x)],\\
    \tilde{G}_{1,0}(E,k_x)&=-\frac{i}{2}[\tilde{G}^\mathrm{A}_{1,0}(E,k_x)-\tilde{G}^\mathrm{R}_{1,0}(E,k_x)],
\end{align}%
where $\tilde{G}^\mathrm{R}_{j_y,j'_y}(E,k_x)$ and $\tilde{G}^\mathrm{A}_{j_y,j'_y}(E,k_x)$ are the retarded and advanced Green's functions, respectively. 
Here, $\mathrm{Tr'_{\uparrow\uparrow}}$ denotes the trace onto only the electron space, and $\bar{t}_{J}$ is defined as
\begin{align}
    \bar{t}_{J}=
    \begin{pmatrix}
        \hat{t}_{J} & 0 \\
        0 & \hat{t}_\mathrm{J}
    \end{pmatrix},
\end{align}%
for the definition of the current operator.
We also define the tunneling conductance with up spin
\begin{align}
    \bar\sigma_{\uparrow}(eV)&=\frac{1}{2\pi}\int^{\pi}_{-\pi}dk_x\sigma_{\uparrow}(eV,k_x) ,
\end{align}%
\begin{align}
    &\sigma_{\uparrow}(eV,k_x)\\
    &=\mathrm{Tr'_{\uparrow\uparrow}}[\tilde{g}_{1}(eV,k_x)+\tilde{g}_{2}(eV,k_x)-\tilde{g}_{3}(eV,k_x)-\tilde{g}_{4}(eV,k_x)],\notag
\end{align}%
and with down-spin
\begin{align}
    \bar\sigma_{\downarrow}(eV)&=\frac{1}{2\pi}\int^{\pi}_{-\pi}dk_x \sigma_{\downarrow}(eV,k_x),
\end{align}%
\begin{align}
    &\sigma_{\downarrow}(eV,k_x)\\
    &=\mathrm{Tr'_{\downarrow\downarrow}}[\tilde{g}_{1}(eV,k_x)+\tilde{g}_{2}(eV,k_x)-\tilde{g}_{3}(eV,k_x)-\tilde{g}_{4}(eV,k_x)],\notag
\end{align}%
where $\mathrm{Tr'_{\uparrow\uparrow}}$ and $\mathrm{Tr'_{\downarrow\downarrow}}$ are the trace onto only $\uparrow\uparrow$ and $\downarrow\downarrow$ in the electron space, respectively.
In the next section, for each pair potential, we calculate $\bar\sigma(eV)$, $\bar\sigma_\uparrow(eV)$, and $\bar\sigma_\downarrow(eV)$ for each momentum $k_x$ and the bias $eV$, and we also show the tunneling conductance. 
Here, we normalized the tunneling conductance by using $\sigma_\mathrm{N}$ with the tunneling conductance in the normal state at zero bias voltage $eV=0$.

\section{Results}

From now on, we show the numerical results of the tunneling conductance in PUM/SC junctions.
We choose the pair potentials as $s$-wave, $d_{x^2-y^2}$-wave, and $d_{xy}$-wave for spin-singlet, and helical $p$-wave, $p_x$-wave, and $p_y$-wave for spin-triplet states in the presence of the time-reversal symmetry.
We also consider spin-singlet chiral $d$ and spin-triplet chiral $p$-wave without the time-reversal symmetry.

\subsection{Spin-singlet superconductor junctions with time-reversal symmetry}

First, we focus on the spin-singlet pair potential.
In Fig.~\ref{fig:k_resolved_singlet}, we plot the momentum-resolved tunneling conductance for $\sigma(eV,k_x)$ and spin-resolved ones $\sigma_\uparrow(eV,k_x)$ and $\sigma_\downarrow(eV,k_x)$ for $s$-wave, $d_{x^2-y^2}$-wave, and $d_{xy}$-wave superconductor junctions in Fig.~\ref{fig:k_resolved_singlet} (a)(d)(g), Fig.~\ref{fig:k_resolved_singlet} (b)(e)(h), and Fig.~\ref{fig:k_resolved_singlet} (c)(f)(i), respectively.
We see the symmetric $\sigma(eV,k_x)$ as a function for both $eV$ and $k_x$ for all spin-singlet states [Figs.~\ref{fig:k_resolved_singlet}(a)(b)(c)].
The obtained $\sigma(eV,k_x)$ are qualitatively the same as those in normal metal/SC junctions: full-gapped state of the $s$-wave [Fig.~\ref{fig:k_resolved_singlet} (a)], the nodal structure of the $d_{x^2-y^2}$-wave [Fig.~\ref{fig:k_resolved_singlet} (b)] and the ZBCP originating from the zero-energy SABSs of the $d_{xy}$-wave [Fig.~\ref{fig:k_resolved_singlet} (c)].
Tunneling conductance with up-spin $\sigma_\uparrow(eV,k_x)$ shown in Figs.~\ref{fig:k_resolved_singlet} (d)(e)(f) and with down spin $\sigma_\downarrow(eV,k_x)$ shown in Figs.~\ref{fig:k_resolved_singlet} (g)(h)(i) are not symmetric for $k_x$ and we find the relation: $\sigma_\uparrow(eV,k_x)=\sigma_\downarrow(eV,-k_x)$.
Then the range of $k_x$ contributing to the conductance process with up-spin is restricted for $-k_{x2}\leq k_x\leq k_{x1}$ with $\sigma_\uparrow(eV,k_x)$ [Figs.~\ref{fig:k_resolved_singlet} (d)(e)(f)], while, with down spin for $-k_{x1}\leq k_x\leq k_{x2}$ with $\sigma_\downarrow(eV,k_x)$ [Figs.~\ref{fig:k_resolved_singlet} (g)(h)(i)].
This originates from only one spin configuration of the Fermi surface in PUMs available for charge transport shown in Fig.~\ref{fig:image} (b): for $-k_{x2}\leq k_x\leq -k_{x1}$ with up-spin and $k_{x1}\leq k_{x}\leq k_{x2}$ with down-spin.

We obtain the tunneling conductance $\bar\sigma(eV)$ after the summing up of $k_x$ in $\sigma(eV,k_x)$ shown in Fig.~\ref{fig:singlet} with $t_J=0$ [Figs.~\ref{fig:singlet} (a)(b)(c)] and $t_J=0.25$ [Figs.~\ref{fig:singlet} (d)(e)(f)].  
Since $\sigma(eV,k_x)$ is symmetric for $eV$, the tunneling conductance $\bar\sigma(eV)$ also becomes symmetric for $eV$ and this feature is the same as those in normal metal/SC junctions at the low transparency limit~\cite{Tanaka95,Tanaka96,Kashiwaya_2000} shown in Figs.~\ref{fig:singlet} (a)(b)(c).
We note that the tunneling conductance for $d_{x^2-y^2}$-wave pair potential at $t_J=0$ shown in Fig.~\ref{fig:singlet} (b) is not V-shape because the nodal structure of $d_{x^2-y^2}$-wave SCs does not appear with $k_{x}$ in the overlapped two Fermi surfaces on the two sides of the junction. 
Indeed, the tunneling conductance in PUM/spin-singlet SC junctions is insensitive to the PUM order.

\subsection{Spin-triplet superconductor junctions with the time-reversal symmetry}
Next, we demonstrate the tunneling conductance of the spin-triplet SC junctions with time-reversal symmetry.
We plot $\sigma(eV,k_x)$, $\sigma_\uparrow(eV,k_x)$, and $\sigma_\downarrow(eV,k_x)$ for helical $p$-wave, $p_x$-wave, and $p_y$-wave SC junctions in Figs.~\ref{fig:k_resolved_triplet} (a)(d)(g), Figs.~\ref{fig:k_resolved_triplet} (b)(e)(h), and Figs.~\ref{fig:k_resolved_triplet} (c)(f)(i), respectively. 
In the case of the helical $p$-wave junction, we obtain the missing of the helical edge states from $-k_{x2}\le k_{x}\le -k_{x1}$ and $k_{x1}\le k_{x}\le k_{x2}$ at $eV\sim 0.8\Delta$ as seen from $\sigma(eV,k_x)$ plotted in Fig.~\ref{fig:k_resolved_triplet} (a).
For the $p_x$-wave SC junctions, because we consider the junction along the $y$-direction, we obtain the nodal structure in the conductance [Fig.~\ref{fig:k_resolved_triplet} (b)] and ZBCP is absent. 
On the other hand, in the $p_y$-wave SC junctions, because we consider the junction along the $y$-direction, we obtain ZBCP caused by the zero-energy ABSs of the $p_y$-wave pairing [Fig.~\ref{fig:k_resolved_triplet} (c)].
When we see the tunneling conductance with up-spin $\sigma_{\uparrow}(eV,k_x)/\sigma_\mathrm{N}$, the range of $k_x$ contributing to the conductance is restricted for $-k_{x2}\le k_x\le k_{x1}$ [Figs.~\ref{fig:k_resolved_triplet} (d)(e)(f)]. 
On the other hand, for tunneling conductance with down spin, $\sigma_{\downarrow}(eV,k_x)/\sigma_\mathrm{N}$ becomes nonzero for $-k_{x1}\le k_x\le k_{x2}$ [Figs.~\ref{fig:k_resolved_triplet} (g)(h)(i)].
We also find the relation: $\sigma_\uparrow(eV,k_x)=\sigma_\downarrow(eV,-k_x)$.

The resulting tunneling conductance $\bar\sigma(eV)$ is plotted in Fig.~\ref{fig:triplet} with $t_J=0$ and $t_J=0.25t$ shown in Figs.~\ref{fig:triplet} (a)(b)(c) and Figs.~\ref{fig:triplet} (d)(e)(f).
For the helical $p$-wave SC junctions, in the absence of the PUM order, the tunneling conductance is symmetric for $eV$ [Fig.~\ref{fig:triplet} (a)].
However, because the momentum resolved tunneling conductance $\sigma(eV,k_x)$ is asymmetric for $eV$ [Fig.~\ref{fig:k_resolved_triplet} (a)], $\bar\sigma(eV)$ shown in green line of Fig.~\ref{fig:triplet} (d) is not symmetric for $eV$.
Then the value of $[\bar\sigma(eV)-\bar\sigma(-eV)]/\sigma_\mathrm{N}$ becomes finite, caused by this asymmetry [brown dotted line in Fig.~\ref{fig:triplet} (d)].
This asymmetry originates from the missing of helical edge states for $-k_{x2}\le k_x\le -k_{x1}$ and $k_{x1}\le k_x\le k_{x2}$ at $eV\sim 0.8\Delta$ [Figs.~\ref{fig:k_resolved_triplet} (d)(g)], and it is caused by the specific spin configuration on the Fermi surface in PUMs.
In $p_x$ and $p_y$-wave SC junctions, the tunneling conductance is symmetric for $eV$ [Fig.~\ref{fig:triplet} (e) and Fig.~\ref{fig:triplet}(f)]. 
These line shapes are the same as those in normal metal/SC junctions shown in Fig.~\ref{fig:triplet} (b) and Fig.~\ref{fig:triplet}(c): V-shape for the $p_x$-wave [Fig.~\ref{fig:triplet} (e)] and ZBCP for the $p_y$-wave SC junctions [Fig.~\ref{fig:triplet} (f)]~\cite{Tanaka95,Tanaka96,Kashiwaya_2000}.
Thus, the line shape of the tunneling conductance in the helical $p$-wave SC junction can be asymmetric for $eV$ in the presence of the PUM order, and in the spin-triplet $p_x$ and $p_y$-wave SC junctions, it is almost independent of the PUM order.

\subsection{Chiral $d$ and chiral $p$-wave superconductor junctions}
In this subsection, we investigate the case of the chiral $d$ and chiral $p$-wave SC junctions without time-reversal symmetry.
As well as the previous subsections, we first show the momentum-resolved tunneling conductance shown in Fig.~\ref{fig:k_resolved_chiral}.
$\sigma(eV,k_x)$ is enhanced at the corresponding chiral edge states inside the energy gap $-\Delta\le eV\le \Delta$ in the chiral $d$ and chiral $p$-wave superconductors [Figs.~\ref{fig:k_resolved_chiral} (a)(b)(c)(d)].
Then $\sigma(eV,k_x)$ in the chiral $d$-wave SC junctions at $t_J=0.25t$ and that of the chiral $p$-wave ones 
at $t_J=0,0.25t$ cross the zero bias voltage [Figs.~\ref{fig:k_resolved_chiral} (b)(c)(d)(f)(g)(h)(j)(k)(l)], while 
$\sigma(eV,k_x)$ in the chiral $d$-wave ones at $t_J=0$ does not cross $eV=0$ [Figs.~\ref{fig:k_resolved_chiral} (a)(e)(i)].
It means that chiral edge states, which can contribute to the conductance at  $eV=0$ 
appears in the chiral $d$-wave SC junctions in the presence of the PUM order.
When we see $\sigma_\uparrow(eV,k_x)/\sigma_\mathrm{N}$ ($\sigma_\downarrow(eV,k_x)/\sigma_\mathrm{N}$), the range of $k_x$ contributing to the conductance is also restricted for $-k_{x2}\le k_x\le k_{x1}$ [Fig.~\ref{fig:k_resolved_chiral} (f) and Fig.~\ref{fig:k_resolved_chiral} (h)] ($-k_{x1}\le k_x\le k_{x2}$) [Fig.~\ref{fig:k_resolved_chiral} (j) and Fig.~\ref{fig:k_resolved_chiral} (l)].
We also find $\sigma_\uparrow(eV,k_x)=\sigma_\downarrow(-eV,-k_x)$.

The tunneling conductance $\bar\sigma(eV)$ shown in Fig.~\ref{fig:chiral} (c) and Fig.~\ref{fig:chiral} (d) at $t_J=0.25t$ is symmetric for $eV$ in the presence of the PUM order. 
As compared with Fig.~\ref{fig:chiral} (a) and Fig.~\ref{fig:chiral} (b) at $t_J=0$, the tunneling conductance 
with up-spin $\bar\sigma_\uparrow(eV)$ and that with down-spin $\bar\sigma_\downarrow(eV)$ is not symmetric for $eV$ inside the energy gap $-\Delta\le eV\le\Delta$.
Owing to the missing of $\sigma_{\uparrow}(eV,k_{x})$ for $k_{x1}\le k_x\le k_{x2}$ [Figs.~\ref{fig:k_resolved_chiral} (f)(h)] and that of $\sigma_{\downarrow}(eV,k_{x})$ for $-k_{x2}\le k_x\le -k_{x1}$ [Figs.~\ref{fig:k_resolved_chiral} (j)(l)], we find that the spin-resolved tunneling conductance has different values for the spin sectors inside the energy gap $-\Delta\le eV\le\Delta$.
This similar behavior is also reported in Ref.\ \cite{BursetPRB2015,BreunigPRB2021}  
It can be applied to spin current in superconducting  spintronics~\cite{Sarma_RevModPhys_2004,Eschrig2015,Linder_2015,Newhorizons_spintronics,Baltz_RevModPhys_2018,HIROHATA2020166711}.
In Fig.~\ref{fig:chiral} (e) and Fig.~\ref{fig:chiral} (f), we plot $\bar\sigma(eV)$ at $eV=0$ as a function of $t_J$.
In general, the magnetic order suppresses the tunneling conductance at $eV=0$ as shown in chiral $p$-wave SC junctions [Fig.~\ref{fig:chiral} (f)], owing to the suppression of the Andreev reflection.
In chiral $d$-wave SC junctions, $\sigma(eV,k_{x})$ cross the zero bias voltage at a finite momentum for $t_J=0.25t$ [Fig.~\ref{fig:k_resolved_chiral} (b)], 
while it does not for $t_J=0$ [Fig.~\ref{fig:k_resolved_chiral} (a)].
Whether the chiral edge states cross at zero energy or not is determined by the size of the Fermi surface in the normal metal or the magnet.
In the present study, the size of the Fermi surface in PUMs, that is, $k_{x2}$, becomes larger with the increase of $t_J$, as shown in Fig.~\ref{fig:kx1_kx2}.
As a result, the tunneling conductance at $eV=0$ is enhanced with the increase of $t_J$ in chiral $d$-wave SC junctions [Fig.~\ref{fig:chiral} (e)].
In the Appendix, we demonstrate the tunneling conductance for the chiral $p$-wave SC junctions with the in-plane \textbf{d}-vector.
The direction of the \textbf{d}-vector does not affect the tunneling conductance in the chiral $p$-wave SC junctions.

\section{Conclusion}\label{sec6}
In this paper, we studied the tunneling conductance in superconducting junctions with PUMs 
using the effective PUM model in Ref.\ \cite{hellenes2024P}.
In spin-singlet and spin-triplet $p_x$ and $p_y$-wave SC junctions, the line shape of the tunneling conductance is not sensitive to the PUM order, even though the range of $k_x$ contributing to the spin-resolved conductance is restricted. 
The tunneling conductance in the helical $p$-wave SC junction becomes asymmetric with respect to $eV$ due to the missing of helical edge states contributing to the conductance. 
It is caused by the spin-split feature of the Fermi surface in PUMs. 
In chiral $d$ and chiral $p$-wave SC junctions, we found that the tunneling conductance is sensitive to the direction of the injected spin because of the time reversal symmetry breaking in SCs.

We have clarified that our obtained results based on the effective model Hamiltonian of PUM with the time-reversal symmetry breaking in PUM/SC junctions are qualitatively similar to those in Ref.\ \cite{maeda2024}, where the exchange coupling appearing in the Hamiltonian for PUMs is described by the odd function of Fermi momentum $\hat{H}_\mathrm{PUM}\propto k$ preserving the time-reversal symmetry.
In our results, we chose the junction along the $y$-direction, where the lattice translation symmetry is preserved. 
As a result, our calculations do not change as compared to the simplified continuum model~\cite{maeda2024}. 
In the present calculation, we chose the chemical potential as $\mu_{p}=-3.85t$ and the Fermi surface shown in Fig.~\ref{fig:image} (b) for PUMs. 
We confirmed that our results are generic by changing the chemical potential of PUMs and SCs.
Indeed, the configuration of the electronic structure described by the simplified Hamiltonian $\hat{H}_\mathrm{PUM}\propto k_y$ \cite{maeda2024,BrekkePRL2024} does not qualitatively deviate from that of the effective model with time-reversal symmetry breaking proposed in Ref.\ \cite{hellenes2024P}, within the calculations of the tunneling conductance.
Therefore, our results are important for the justification of the phenomenological analysis in the simplified continuum model in Ref.\ \cite{maeda2024}.
While, along the $x$-direction, because the system does not have the translational symmetry along the direction which is normal to the interface, $[C_2||t]$-symmetry~\cite{hellenes2024P} is broken.
Thus, the effect of the time-reversal symmetry breaking can be expected in junctions along the $x$-direction.

\begin{figure}[t!]
    \centering
    \includegraphics[width=8.5cm]{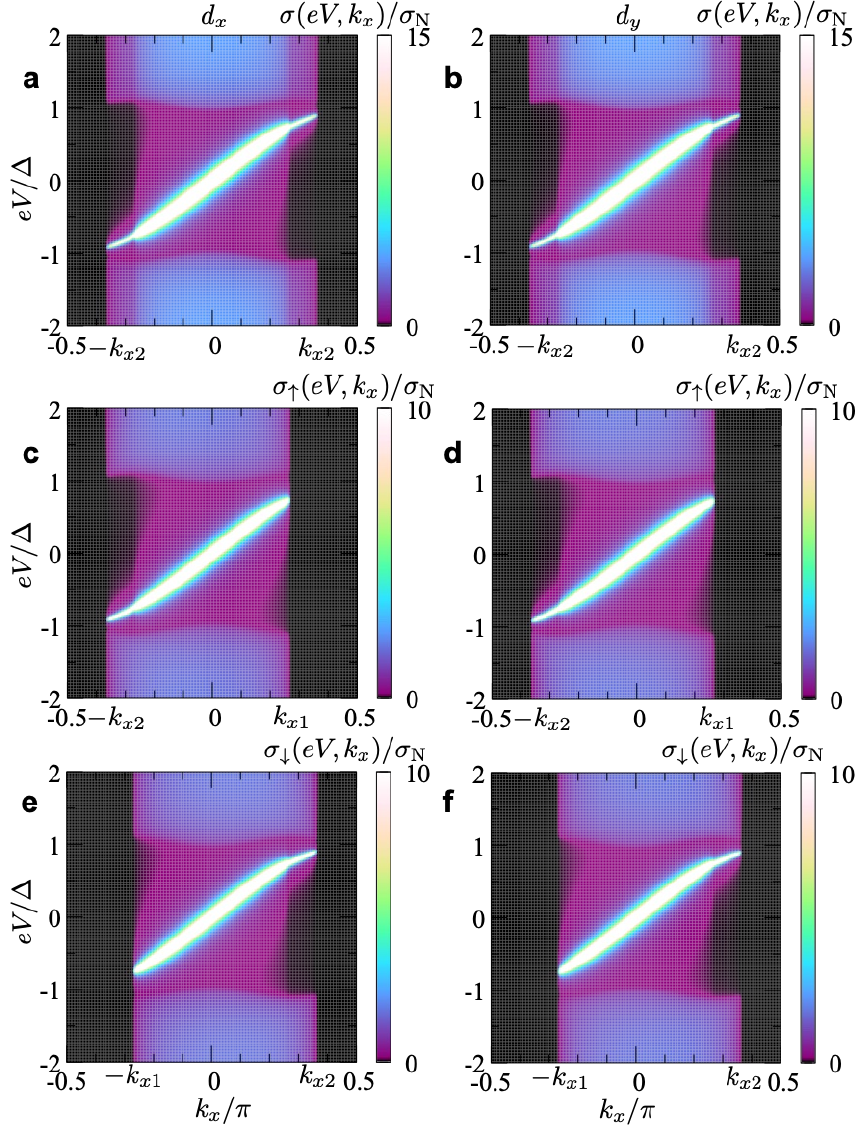}
    \caption{(a)(b) Momentum-resolved tunneling conductance $\sigma(eV,k_x)$ chiral $p$-wave SC junctions, where the \textbf{d}-vector is along (a) $x$ and (b) $y$-direction.
    Here, $\sigma_\mathrm{N}$ is the tunneling conductance in the normal state at $eV=0$.
    (c)(d) Momentum-resolved tunneling conductance with up-spin $\sigma_{\uparrow}(eV,k_x)$ for chiral $p$-wave SC junctions, where the \textbf{d}-vector is along (c) $x$ and (d) $y$-direction.
    (e)(f) Momentum-resolved tunneling conductance with down-spin $\sigma_{\downarrow}(eV,k_x)$ for chiral $p$-wave SC junctions, where the \textbf{d}-vector is along (e) $x$ and (f) $y$-direction. 
    We choose the parameters as $t_J=0.25t$ $U_\mathrm{b}=5t$, $\Delta=0.001t$, and $\delta=0.01\Delta$.}
    \label{afig:k_resolved_chiral_xy}
\end{figure}%
\begin{figure}[t!]
    \centering
    \includegraphics[width=8.5cm]{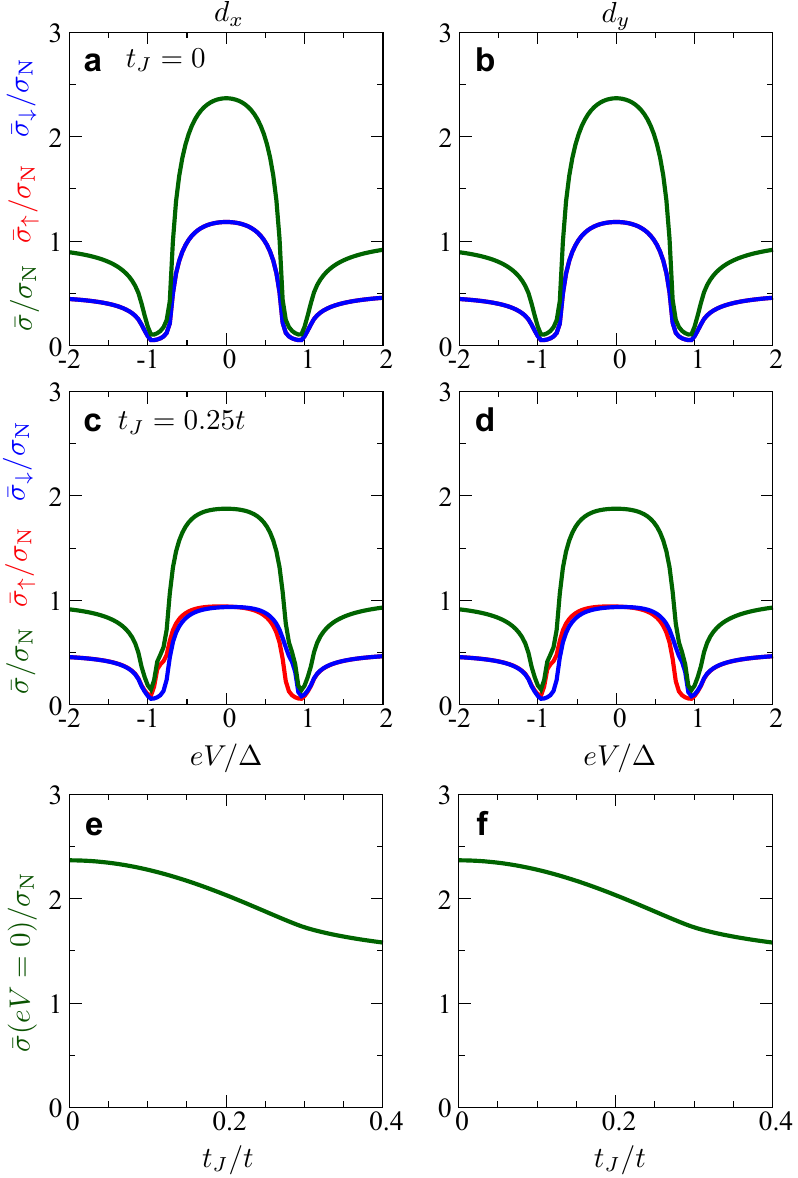}
    \caption{Tunneling conductance $\bar\sigma(eV)$ for chiral $p$-wave at (a)(b) $t_J=0$ and (c)(d) $t_J=0.25t$.
    This tunneling conductance is normalized by its value in the normal state at $eV=0$: $\sigma_\mathrm{N}$.
    In (a)(c), the \textbf{d}-vector is along the $x$-direction, and in (b)(d), along the $y$-direction.
    Green, red and blue colors indicate $\bar\sigma(eV)/\sigma_\mathrm{N}$, $\bar\sigma_\uparrow(eV)/\sigma_\mathrm{N}$, and $\bar\sigma_\downarrow(eV)/\sigma_\mathrm{N}$, respectively.
    (e)(f) Tunneling conductance at $eV=0$ as a function of the strength of the PUM order $t_J$ for the chiral $p$-wave SC junctions whose \textbf{d}-vector is along (e) $x$ and (f) $y$-direction.
    We choose the parameters as $U_\mathrm{b}=5t$, $\Delta=0.001t$, and $\delta=0.01\Delta$.}
    \label{afig:chiral_xy}
\end{figure}%

In our work, we set the isotropic tunneling Hamiltonian at the interface.
For future perspective, to enhance the effect of time-reversal symmetry breaking, we can focus on the anisotropic tunneling hopping at the interface and another direction of junctions ($x$-direction). 
In addition, the Josephson junctions with PUM without time-reversal symmetry need to be discussed.
In Ref.\ \cite{fukaya2024}, the Josephson current is calculated by using the simplified PUM Hamiltonian 
with time reversal symmetry \cite{maeda2024}.
Indeed, the role of time-reversal symmetry breaking in PUM should also be clarified in Josephson junctions with PUMs.
Also, the analysis of the pair amplitude, including the odd-frequency pairing 
\cite{tanaka12}, is needed beyond previous studies 
based on the simplified Hamiltonian \cite{fukaya2024,MaedaPRB2025}.

\begin{acknowledgements}
 Y.\ F.\ and K.\ Y.\ acknowledge financial support from the Sumitomo Foundation.
 Y.\ F.\ also acknowledges the numerical calculation support from Okayama University.    
 Y.\ T.\ acknowledges financial support from JSPS with Grants-in-Aid for Scientific Research (KAKENHI Grants Nos. 23K17668, 24K00583, 24K00556, 24K00578, 25H00609, and 25H00613).  
 We thank S.\ Ikegaya, J.\ Cayao, B.\ Lu, and P.\ Sukhachov for the valuable discussions.
\end{acknowledgements}


\appendix

\section{Chiral $p$-wave superconductor junctions with an in-plane \textbf{d}-vector}

In the Appendix, we show the tunneling conductance for chiral $p$-wave SC junctions with in-plane \textbf{d}-vector.
We define the pair potentials as
\begin{align}
    d_{x}(\bm{k})&=\Delta[\sin{k_x}+i\sin{k_y}],\\
    d_{y}(\bm{k})&=0, \hspace{2mm}d_{z}(\bm{k})=0,\notag 
\end{align}%
and 
\begin{align}
    d_{x}(\bm{k})&=0,\notag\\
    d_{y}(\bm{k})&=\Delta[\sin{k_x}+i\sin{k_y}],\\
    d_{z}(\bm{k})&=0. \notag
\end{align}%

We plot the momentum-resolved tunneling conductance and each spin component for $eV$ and $k_x$ in Fig.~\ref{afig:k_resolved_chiral_xy} with the \textbf{d}-vectors along the $x$ and $y$-direction.
The structure of $\sigma(eV,k_x)$ is independent of the direction of the \textbf{d}-vector shown in Fig.~\ref{afig:k_resolved_chiral_xy} (a) and Fig.~\ref{afig:k_resolved_chiral_xy} (b), as well as $\sigma_\uparrow(eV,k_x)$ [Fig.~\ref{afig:k_resolved_chiral_xy} (c) and Fig.~\ref{afig:k_resolved_chiral_xy} (d)], and $\sigma_\downarrow(eV,k_x)$ [Fig.~\ref{afig:k_resolved_chiral_xy} (e) and Fig.~\ref{afig:k_resolved_chiral_xy} (f)].
Compared with Figs.~\ref{fig:k_resolved_chiral} (d)(h)(l) and Fig.~\ref{afig:k_resolved_chiral_xy}, we obtain almost the same results for the \textbf{d}-vector along the $z$-direction.
As well as the tunneling conductance $\bar\sigma(eV)$, we obtain almost the same results at both $t_J=0$ and $t_J=0.25t$ [Fig.~\ref{afig:chiral_xy}].
Indeed, the direction of the \textbf{d}-vector does not affect the tunneling conductance in the chiral $p$-wave SC junctions.


\bibliography{biblio}

\end{document}